\documentclass[particles,article,accept,moreauthors,pdftex,10pt,a4paper]{mdpi} 

\firstpage{1} 
\articlenumber{x}
\doinum{10.3390/------}
\pubvolume{2}
\pubyear{2018}
\copyrightyear{2018}
\history{Received: 11 April 2018; Accepted: 1 May 2018; Published: date}

\pdfoutput=1

 \theoremstyle{mdpi}
 \newcounter{thm}
 \newcounter{ex}
 \newcounter{re}
 \theoremstyle{mdpidefinition}
\usepackage[T1]{fontenc}
\usepackage{graphicx}
\usepackage{bm}
\usepackage{amssymb}
\usepackage{amsmath,stmaryrd}
\usepackage{amsfonts}
\usepackage{dcolumn}
\usepackage{natbib}
\usepackage{color}
\usepackage{calc}
\usepackage{longtable}
\usepackage{pdflscape}
\usepackage[normalem]{ulem}
\usepackage{soul}

\newcommand{\raiseentry}[1]{\smash{\raise 0.7 em \hbox{#1}}}

\newcommand{\lowentry}[1]{\smash{\lower 1.5 ex \hbox{#1}}}

\newcommand{\up}{\upsilon}
\newcommand{\be}{{\bar\epsilon}}

\newcommand{\eg}{{e.g.}}

\def\apj{Astrophys. J.}
\def\apjl{Astrophys. J. Lett.}

\def\aap{Astron. Astrophys. }

\def\rmp{Rev.\ Mod.\ Phys.\ }

\newenvironment{equationarray*}
{\arraycolsep 0.14 em
\begin{eqnarray*}}
{\end{eqnarray*}}

\usepackage{booktabs} 
\usepackage{multirow}
\usepackage{soul} 
\usepackage{microtype}

\Title{Turbulence Generation by Shock-Acoustic-Wave Interaction in Core-Collapse Supernovae}

\Author{{Ernazar Abdikamalov}$^{1,}$*, {C\'esar} Huete $^{2}$, {Ayan} Nussupbekov $^{1}$ and {Shapagat} Berdibek $^{1}$}  
\AuthorNames{E. Abdikamalov, C. Huete, A. Nussupbekov, and S. Berdibek}

\address{%
$^{1}$ \quad Department of Physics, School of Science and Technology, Nazarbayev University, Astana 010000, Kazakhstan; ayan.nussupbekov@nu.edu.kz (A.N.);  shapagat.berdibek@nu.edu.kz (S.B.)\\
$^{2}$ \quad Fluid Mechanics Group, Escuela Polit\'ecnica Superior, Universidad Carlos III de Madrid, 28911, Legan\'es, Spain; chuete@ing.uc3m.es}  

\corres{Correspondence: ernazar.abdikamalov@nu.edu.kz}

\abstract{Convective instabilities in the advanced stages of nuclear shell burning can play an important role in neutrino-driven supernova explosions. In our previous work, we studied the interaction of vorticity and entropy waves with the supernova shock using a linear perturbations theory. In this paper, we extend our work by studying the effect of acoustic waves. As the acoustic waves cross the shock, the perturbed shock induces a field of entropy and vorticity waves in the post-shock flow. We find that, even when the upstream flow is assumed to be dominated by sonic perturbations, the shock-generated vorticity waves contain most of the turbulent kinetic energy in the post-shock region, while the entropy waves produced behind the shock are responsible for most of the density perturbations. The entropy perturbations are expected to become buoyant as a response to the gravity force and then generate additional turbulence in the post-shock region. This leads to a  modest reduction of the critical neutrino luminosity necessary for producing an explosion, which we estimate to be {less than $\sim5\%$}.}

\keyword{hydrodynamics, shock waves, turbulence, supernovae: general}

\begin{document}

\section{Introduction}
\label{section:introduction}

Core-collapse supernova (CCSN) explosions are multi-dimensional phenomena \cite{fryxell:91, herant:95, bhf:95, janka:96, moesta:14b, boggs:15, mao:15, wongwathanarat:15, mueller:16b, katsuda:18}. It has long been recognized that hydrodynamics instabilities such as neutrino-driven convection and standing accretion shock instability play crucial roles in launching CCSN explosions (e.g., \cite{blondin:03, foglizzo:06, foglizzo:07, hanke:12, hanke:13, janka:12b, dolence:13, murphy:13, burrows:13a, takiwaki:14, ott:13neutrino, ott:13a, abdikamalov:15, radice:15a, melson:15a, melson:15b, lentz:15, fernandez:14, fernandez:15a, foglizzo:15, cardall:15, radice:16a, bruenn:16, roberts:16c, kuroda:17, ott:18, takiwaki:18, kazeroni:18, radice:18}). Recently, it has become clear that the convective instabilities that arise in \textls[-5]{the nuclear burning shells may also contribute to the explosion \cite{couch:13d, couch:15a, mueller:15, mueller:16}. As the stellar core begins its collapse, convective shell perturbations accrete towards the center, a journey during which they undergo significant amplification due to the converging geometry of the flow \cite{kovalenko:98,lai:00,takahashi:14}. The supernova shock wave encounters these perturbations within $\sim$1 {s} after formation. The interaction of the perturbations with the shock generates additional turbulence in the post-shock region, creating a more favorable condition for energizing the shock~\cite{mueller:17,abdikamalov:16,huete:18}.} The~perturbations originating from the oxygen and, to a lesser extent, silicon burning shells were found to be particularly important \cite{collins:18}. 

In our previous work \cite{abdikamalov:16,huete:18}, we investigated the physical mechanism of how convective perturbations interact with the supernova shock using a linear perturbation theory known as the linear interaction analysis (LIA) (e.g., \cite{ribner:53,chang:57,mahesh:96,wouchuk:09, huete:11, huete:12}). We used an idealized setup, in which the mean flow is assumed to be uniform, while the perturbations are modeled as planar sinusoidal waves. The linearized jump conditions at the shock provide boundary conditions for the linear Euler equations in the post-shock region, the solution of which yields the flow parameters in that region. The simplicity of such a setup allows us to obtain a unique insight into the physics of the shock--turbulence interaction. 

We considered entropy and vorticity perturbations in the accretion flow, which represent two components of a generic three-component hydrodynamic weak turbulence, the third being acoustic waves, as demonstrated by \textls[-5]{\citep{kovasznay:53}. When any of these perturbations encounters the shock, they deform the shock and generate a field of acoustic, vorticity, and entropy waves in the post-shock region. The~entropy waves become buoyant and generate additional convection in the post-shock region~\cite{mueller:16,mueller:17},} which~then lowers the critical neutrino luminosity necessary for driving the explosion. We estimated the reduction to be $\sim$17--24\%, which is in agreement with the predictions from 3D neutrino-hydrodynamics simulations \cite{mueller:17}. 

The goal of the present work is to extend our previous work \cite{abdikamalov:16,huete:18} by studying the effect of the third component of the turbulent field: the acoustic perturbations. While the convective motion in nuclear burning shells can in principle emit acoustic waves, the efficiency of this process is rather low due to subsonic nature of the turbulence in nuclear burning shells \cite{lighthill:52,landau:59}. Hence, these acoustic waves have a negligible effect on the explosion condition of CCSNe. On the other hand, the accretion of entropy and vorticity waves during stellar collapse can generate significant acoustic waves \cite{kovalenko:98, lai:00, foglizzo:00, foglizzo:01}. In~this paper, we investigate the effect of these waves on the shock dynamics in CCSNe. 

The rest of this paper is organized as following. In Section \ref{sec:lia}, we describe our method. In Section~\ref{sec:results}, we present the result of our investigation of the shock--acoustic--wave interaction. We estimate the strength of vorticity, entropy, and acoustic waves generated in the post-shock region. In Section~\ref{sec:implications}, we~estimate the impact of these perturbations on the explosion condition of CCSNe. Finally, in Section~\ref{sec:summary}, we summarize our results and provide our conclusions.  

\section{Method and Setup}
\label{sec:lia}

As in \cite{abdikamalov:16}, we model the unperturbed flow as a uniform and one-dimensional flow along axis $x$, while unperturbed shock is assumed to be a planar discontinuity perpendicular to $x$. The mean flow parameters on the two sides of the shock are related to each other via the Rankine--Hugoniot conditions~\cite{landau:59}, while the perturbations' parameters are related via the linearized version of the jump conditions (e.g., \cite{moore:54,sagaut:08}). This yields boundary conditions for the linearized Euler equations behind the shock, the solution of which, along with the isolated-shock assumption, completely determines the post-shock flow and the shock dynamics in terms of the upstream flow parameters. We denote the mean velocity, density, pressure, temperature, and Mach number as $U$, $\bar{\rho}$, $\bar{p}$, $\bar{T}$, and $\mathcal{M}$, respectively, while the perturbations in the $x$- and $y$-components of velocity, density, pressure, and temperature are denoted by $u'$, $\up'$ $\rho'$, $p'$, $T'$, respectively. The upstream and downstream values of these quantities will be denoted with subscripts 1 and 2, respectively. The speed of sound, which determines the propagation velocity of the sonic perturbations relative to the fluid particles, is given by $\bar{a}^2=\gamma \bar{p}/\bar{\rho}$, as~dictated by the perfect-gas equation of state. Our method is a direct extension of the models of Moore~\cite{moore:54} and \citet{mahesh:95} to shocks with endothermic nuclear dissociation, which is an important property of shocks in CCSNe (e.g., \cite{bethe:90, huete:18}). As in \cite{abdikamalov:16}, this model assumes that hydrodynamic time response is much longer than the characteristic time of the nuclear-dissociating process. That is, the shock front and the layer where the dissociation of heavy nuclei into light nuclei takes place are conjointly taken as a single discontinuity. In addition, it is assumed that the intensity of the acoustic perturbations upstream is smaller than the uncertainty inherent to the nuclear dissociation model. Then, the energy involved in the dissociation process can be taken unaltered in the perturbation analysis in the first approximation, without significant loss of accuracy.

\begin{figure}[H]
\begin{center}
 \includegraphics[angle=0,width=0.49\columnwidth,clip=false]{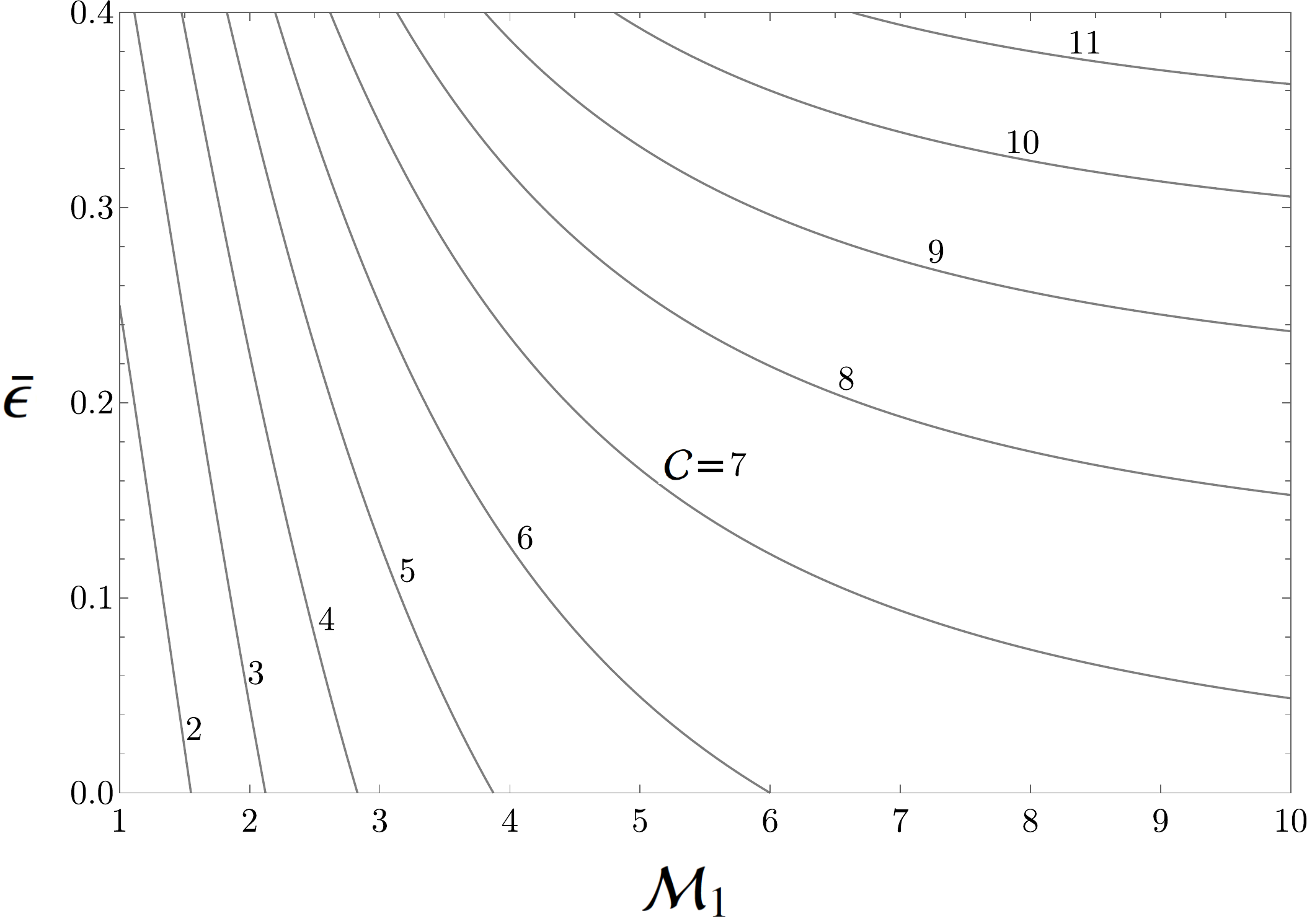} \includegraphics[angle=0,width=0.49\columnwidth,clip=false]{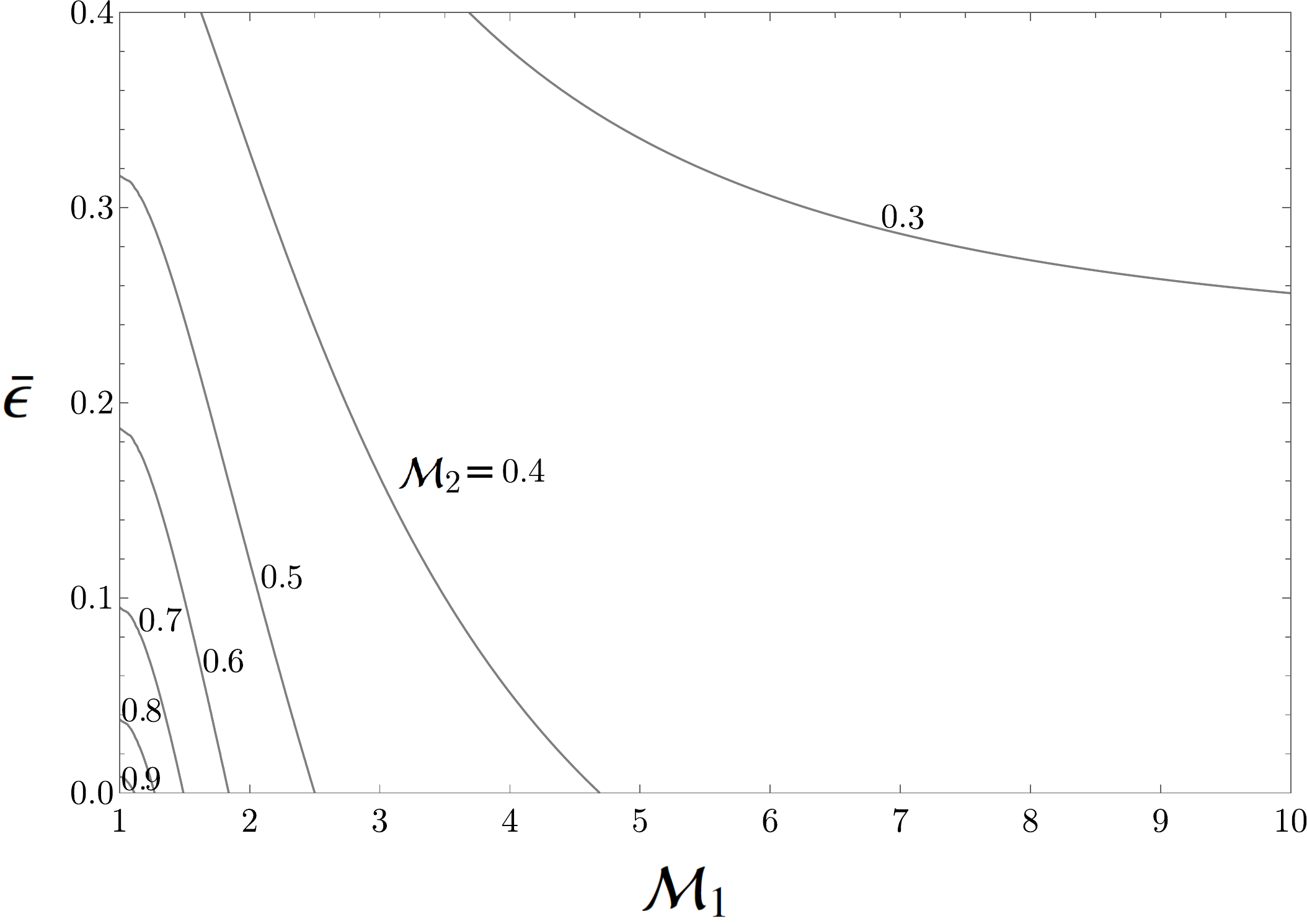}
  \caption{Mass-compression ratio ({\bf left panel}) and post-shock Mach number ({\bf right panel}) as a function of nuclear-dissociation parameter $\be$ and upstream Mach number ${\cal M}_1$.
  \label{fig:CandM2}}
\end{center}
\end{figure}

We model the stellar fluid with an ideal gas equation of state with $\gamma=4/3$. The mean flow parameters are chosen to approximate the situation in CCSNe by requiring vanishing Bernoulli parameter above the shock, as described in \citet{abdikamalov:16}. Due to this condition, we have only two free parameters to characterize the mean flow. We choose the upstream Mach number ${\cal M}_1$ and the nuclear dissociation energy as our free parameters. We express the latter in terms of the dimensionless nuclear dissociation parameter $\be$, which represents the ratio of the specific nuclear dissociation energy to the local free-fall specific kinetic energy \cite{fernandez:09a,fernandez:09b}. It ranges from $0$, which corresponds to the limit of no nuclear dissociation, to $\sim$0.4, which corresponds to strong nuclear dissociation \cite{huete:18}. The upstream Mach number ${\cal M}_1$ typically ranges between $\sim$5 and $\sim$15 for accretion shocks prior to explosion when the shock radius is $\lesssim$300 {km} \cite{huete:18}. The compression factor ${\cal C}$ at the shock ($=\!\bar{\rho}_2/\bar{\rho}_1$),
\begin{eqnarray}
{\mathcal C} &=& \frac{\left(\gamma+1\right){\cal M}_1^2}{\gamma {\cal M}_1^2+1 - \sqrt{\left({\cal M}_1^2-1\right)^2+\be (\gamma+1){\cal M}_1^2 \left[2+ (\gamma-1){\cal M}_1^2\right]}},
\label{C}
\end{eqnarray}
and the post-shock Mach number ${\cal M}_2$ ($=\!U_2/\bar{a}_2$),
\begin{eqnarray}
{\cal M}_2 &=& \left[\frac{\gamma {\cal M}_1^2+1 - \sqrt{\left({\cal M}_1^2-1\right)^2+\be (\gamma+1){\cal M}_1^2 \left[2+ (\gamma-1){\cal M}_1^2\right]}}{\gamma {\cal M}_1^2+1 +\gamma\sqrt{\left({\cal M}_1^2-1\right)^2+\be (\gamma+1){\cal M}_1^2 \left[2+ (\gamma-1){\cal M}_1^2\right]}}\right]^{1/2},\label{M2}
\end{eqnarray}
are depicted in {Figure}~\ref{fig:CandM2} as a function of $\be$ and ${\cal M}_1$. It is found that the post-shock Mach number is very sensitive to the nuclear dissociation parameter in the weak-shock domain and that sonic conditions downstream cannot be achieved whenever $\be>0$, a characteristic of endothermic supersonic propagating waves. In the strong-shock limit, a more realistic limit in the CCSNe context, the above relationships \eqref{C} and \eqref{M2} reduce to ${\mathcal C}=7/\left(4-\sqrt{9+7\be}\right)$ and ${\cal M}_2= \left[3\sqrt{9+7\be}-3(3+\be)\right]^{1/2}/(2\sqrt{\be})$  for $\gamma=4/3$. 

Jump conditions for other quantities that might be of interest are easily written in terms of the functions \eqref{C} and \eqref{M2}, and the upstream Mach number ${\cal M}_1$. They include velocity $U_2/U_1= \mathcal C^{-1}$, speed~of sound and temperature $\bar{a}_2/\bar{a}_1=\bar{T}_2^2/\bar{T}_1^2= {\cal M}_1/({\mathcal C} {\cal M}_2)$, and pressure $\bar{p}_2/\bar{p}_1= {\cal M}_1^2/({\mathcal C} {\cal M}_2^2)$.

\begin{figure}[H]
\begin{center}
 \includegraphics[angle=0,width=0.48\columnwidth,clip=false]{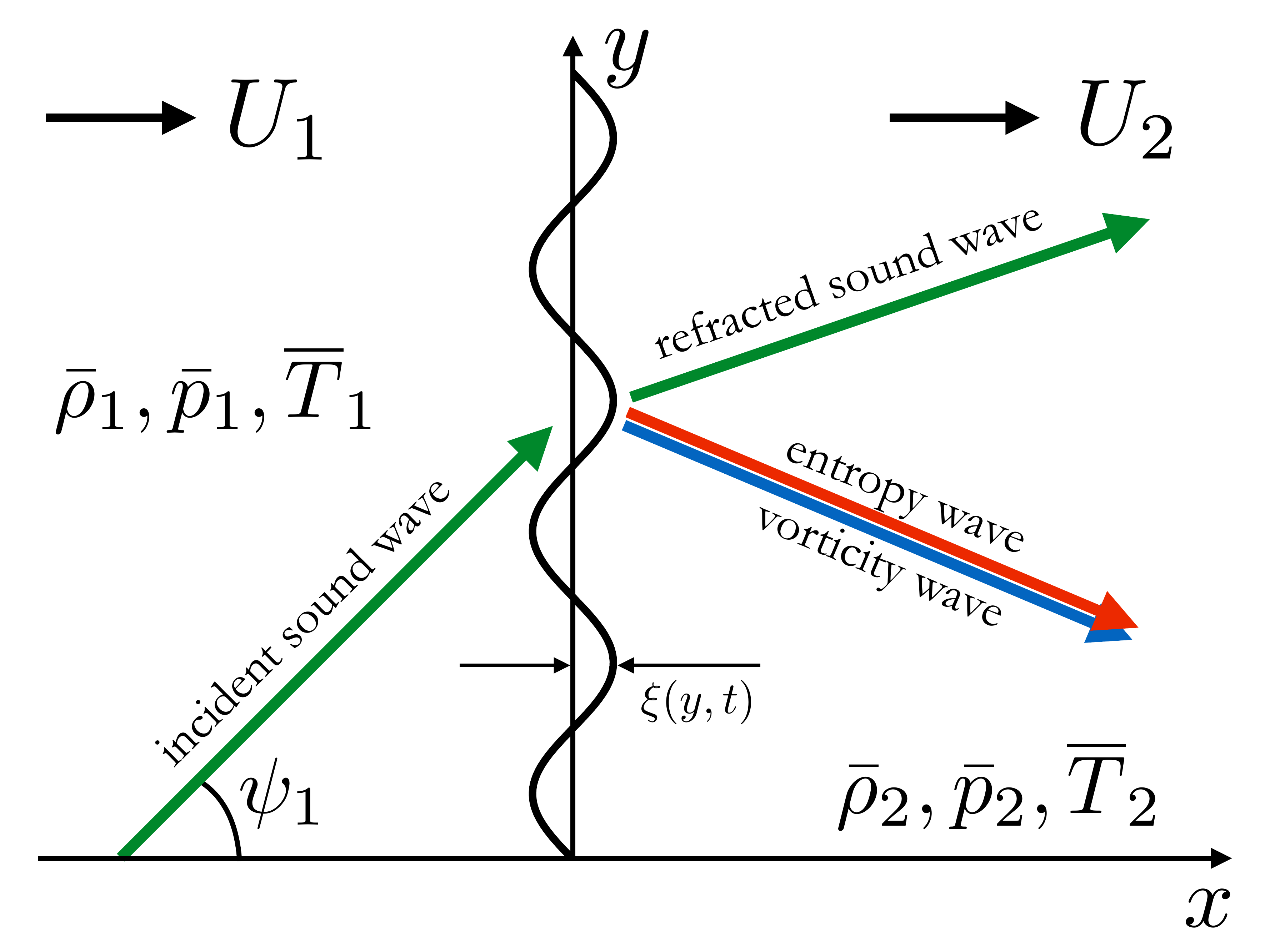}
  \caption{Schematic depiction of the interaction of an acoustic wave with a shock wave. The~unperturbed shock is located at $x=0$ and the mean flow is in the direction of the positive $x$-axis. The pre-shock unperturbed flow is described by velocity $U_1$, pressure $\bar p_1$, density $\bar{\rho}_1$, and~temperature $\overline T_1$, while the post-shock counterparts are $U_2$, $\bar{\rho}_2$, $\bar p_2$, and $\overline T_2$. The deformation of the shock as a response to acoustic perturbations is described by function $\xi(y,t)$. The perturbed shock generates a field of entropy and vorticity waves in the post-shock region. 
  \label{fig:shocklia}}
\end{center}
\end{figure}

To characterize the interaction with an isotropic field of acoustic perturbations, it is convenient to begin the analysis considering a mono-chromatic field of sound waves, as sketched in {Figure} \ref{fig:shocklia}. Then, in a reference frame $x_1-y$ comoving with the mean flow, an incident sinusoidal planar acoustic wave with amplitude $A_p$ and wavelength $\lambda_1$ can be expressed as (e.g., \cite{moore:54,mahesh:95,huete:12})
\begin{equation}
\label{eq:p1}
\frac{p'_1}{\bar{p}_1} = A_p \exp \left( 2 \pi i \frac{mx_1-ly-\bar{a}_1t}{\lambda_1} \right), 
\end{equation}
with the rest of the thermodynamical perturbations being obtained through the isentropic and irrotational conditions that define the upstream potential flow through the Euler equations, namely
\begin{equation}
\bar{\rho}_1\frac{\partial u_1'}{\partial t} =-\frac{\partial p_1'}{\partial x},\quad  \bar{\rho}_1\frac{\partial \up_1'}{\partial t} =-\frac{\partial p_1'}{\partial y},\quad \bar{a}_1^2 \frac{\partial \rho_1'}{\partial t} = \frac{\partial p_1'}{\partial t}.
\end{equation}

Here, $t$ is time, $m=\cos\psi_1$, and $l=\sin\psi_1$, where $\psi_1$ is the angle between the propagation direction of the acoustic wave and the $x_1$-axis. Thus, the incident acoustic waves are completely determined via parameters $A_p$ and $\psi_1$. For a given incident acoustic wave perturbation, due to the linearity of the method, the amplitude of post-shock perturbations pressure, density, and velocity are directly proportional to $A_p$. 

An estimate of the upper limit of perturbations amplitude $A_p$ can be inferred from the results of the 3D neutrino-hydrodynamics simulations of \citet{mueller:17}, who observe perturbations of density with relative amplitude of $\sim$0.1 immediately before the shock. These perturbations are a result of a combinations of entropy, vorticity, and acoustic waves. For the sake of estimating an upper limit for the amplitude of acoustic waves, let us assume that all of the density fluctuations are due to acoustic waves. In this case, the pressure perturbation amplitude $A_p$ can at most be $\sim$0.1$\gamma$, where $\gamma$ is the adiabatic index of the equation of state. The angular distribution of incident acoustic perturbations is not known currently. In our work, for simplicity, we assume isotropic distribution. 

\section{Results}
\label{sec:results}

\subsection{Dependence on Incidence Angle}
\label{sec:incident_angle}

Within the framework of the LIA, when an acoustic wave of form (\ref{eq:p1}) hits a planar shock wave, the latter responds by deforming into a form of planar wave propagating in the $y$-direction \cite{huete:12}. This~process generates vorticity and entropy waves in the post-shock flow, which are then advected by the flow in the downstream direction, as depicted schematically in Figure~\ref{fig:shocklia}. Depending on the incidence angle $\psi_1$, two distinct solutions are possible for acoustic waves in the post-shock region. If~$0 \le \psi_1 \le \psi_\mathrm{cl}$ or $\psi_\mathrm{cu} < \psi_1 \le \pi$, the solution in the post-shock region represents freely propagating planar sound waves. Physically, this solution corresponds to refraction of a sound waves through the shock. The critical angles $\psi_\mathrm{cl}$ and $\psi_\mathrm{cu}$ are the roots of equation \citep{moore:54}
\begin{eqnarray}
\label{eq:psic_eq}
\left(\frac{\bar{a}_2}{U_1}\right)^2 - \left(\frac{U_2}{U_1}\right)^2 = \left( \cot \psi_\mathrm{c} + \frac{\bar{a}_1}{U_1} \csc \psi_\mathrm{c} \right)^2.
\end{eqnarray}

On the other hand, if $\psi_\mathrm{cl} < \psi_1 < \psi_\mathrm{cu}$, the solution represents an exponentially damping acoustic wave.~The former regime is referred to as the {\it propagating} regime, while the latter is called the {\it non-propagating} regime. In both of these regimes, the entropy and vorticity waves propagate without~decaying. 

\begin{figure}[H]
\begin{center}
\includegraphics[angle=0,width=0.48\columnwidth, clip=false]{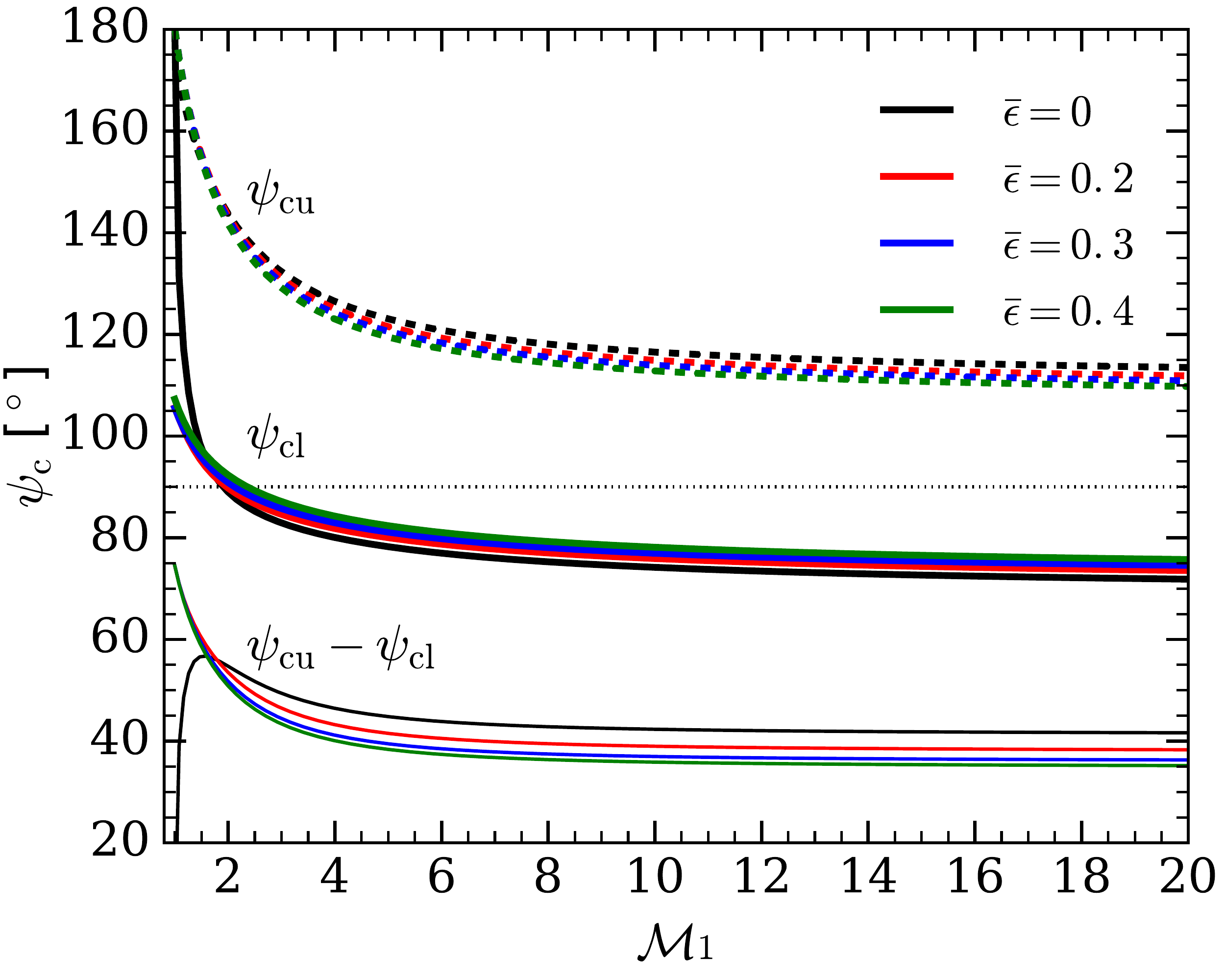}
  \caption{The critical angles $\psi_\mathrm{cl}$ (thick solid lines) and $\psi_\mathrm{cu}$ (dashed lines) versus pre-shock Mach number for four different values of the nuclear dissociation parameter $\be$: $0$, $0.2$, $0.3$, and $0.4$. The thin solid lines depict the difference $\psi_\mathrm{cu}-\psi_\mathrm{cl}$ as a function of the upstream Mach number for the same four values of $\be$. The difference $\psi_\mathrm{cu}-\psi_\mathrm{cl}$ is smaller for larger compression at the shock. Therefore, $\psi_\mathrm{cu}-\psi_\mathrm{cl}$ decreases both with increasing $\be$ and ${\cal M}_1$ for ${\cal M}_1 \gtrsim 2$. 
  \label{fig:psic}}
\end{center}
\end{figure}

Figure~\ref{fig:psic} shows the critical angles $\psi_\mathrm{cl}$ and $\psi_\mathrm{cu}$ as a function of upstream Mach number ${\cal M}_1$ for four values of the nuclear dissociation parameter $\bar \epsilon$: $0$, $0.2$, $0.3$, and $0.4$. In the limit ${\cal M}_1 \rightarrow 1$ and $\be \rightarrow 0$, both $\psi_\mathrm{cl}$ and $\psi_\mathrm{cu}$ equal $180^\circ$, which means that any incoming acoustic waves can propagate into the post-shock region. For $1 < {\cal M}_1 \lesssim 5$, both angles decrease fast with increasing ${\cal M}_1$, reaching, \eg, $121.5^\circ$ and $80^\circ$ at ${\cal M}_1=5$ for $\bar\epsilon=0.2$, after which both angles slowly approach their asymptotic (${\cal M}_1\rightarrow \infty$) values of $109^\circ$ and $71^\circ$. This is a generic property of \textls[-5]{stationary shocks, i.e., parameters of stationary shocks depend weakly on upstream Mach number for ${\cal M}_1 \gtrsim 5$. Note that the asymptotic values of $\psi_\mathrm{cl}$ and $\psi_\mathrm{cu}$ are always symmetric with respect to the $\psi_1=\pi/2$ line (shown with horizontal dashed line), irrespective of the value of $\bar\epsilon$. The width of the non-propagative region, defined as difference $\psi_\mathrm{cu}-\psi_\mathrm{cl}$, decreases with both $\be$ and ${\cal M}_1$. This is a manifestation of the property that $\psi_\mathrm{cu}-\psi_\mathrm{cl}$ decreases with increasing shock compression, which one can achieve by increasing either $\be$ or ${\cal M}_1$. }

\subsection{Interaction with an Isotropic Field of Acoustic Waves}

\begin{figure}[H]
\begin{center}
\includegraphics[angle=0,width=0.45\columnwidth, clip=false]{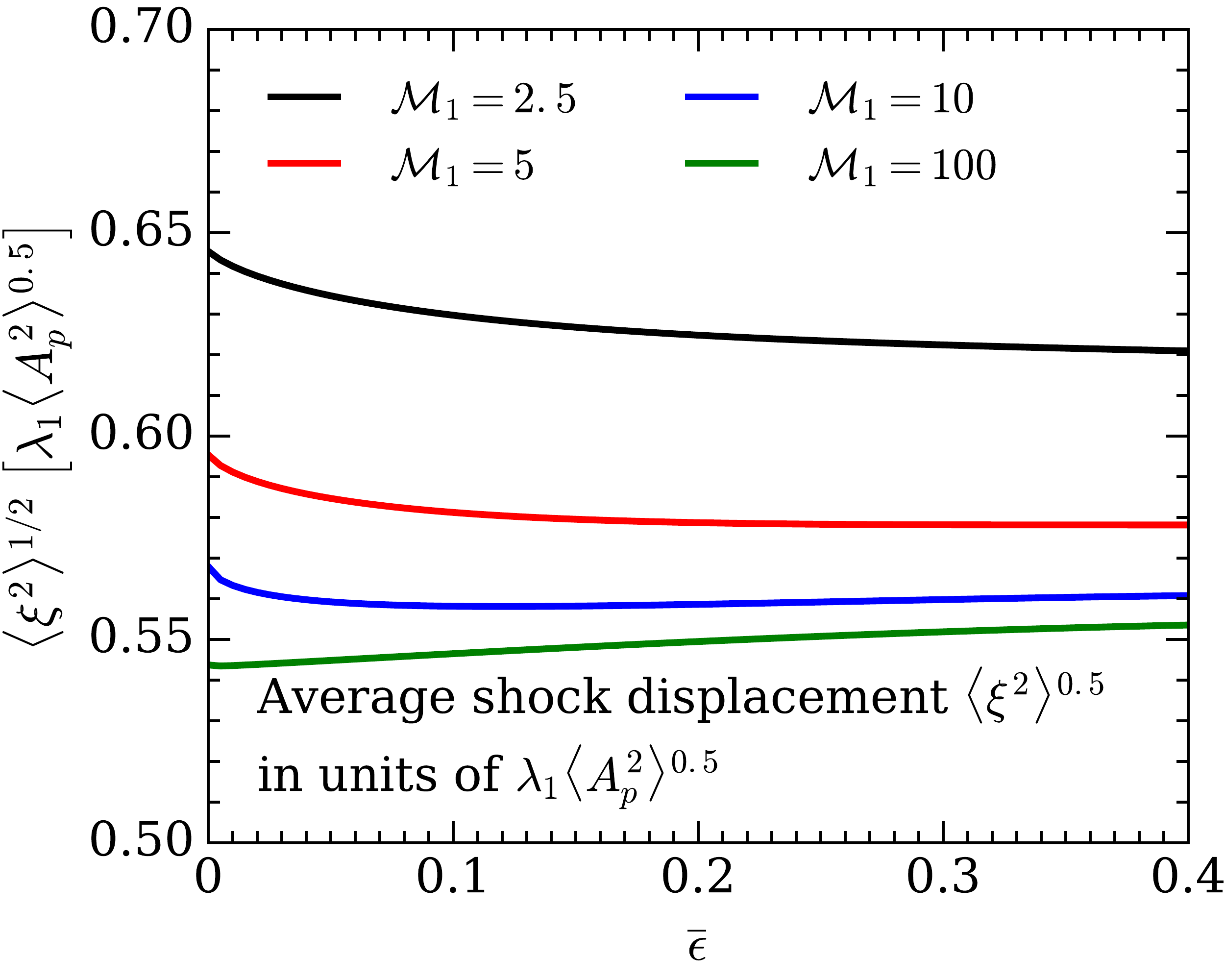}
\includegraphics[angle=0,width=0.45\columnwidth, clip=false]{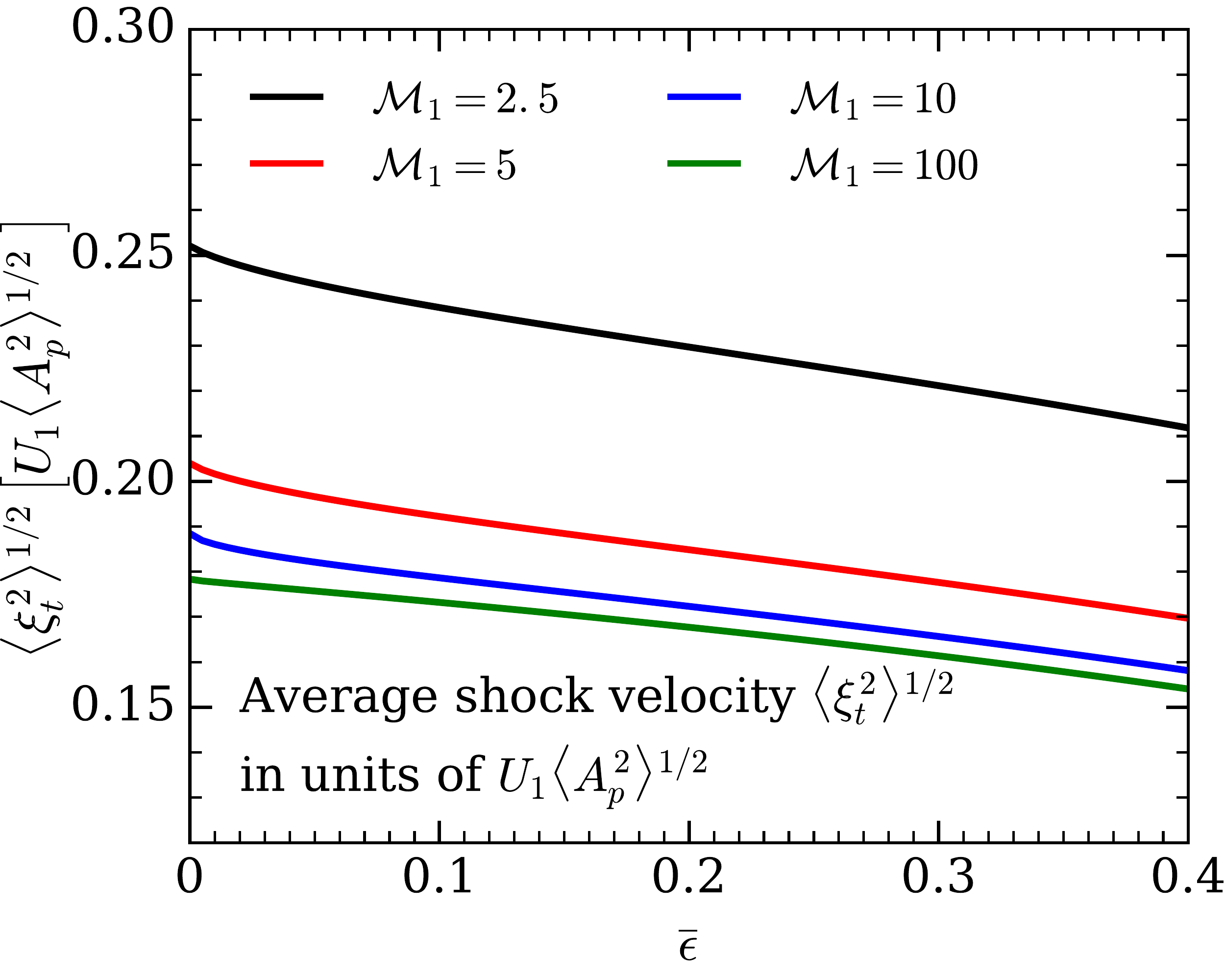}
  \caption{RMS shock displacement ({\bf left panel}) and velocity ({\bf right panel}) as a function of nuclear dissociation parameter for various values of upstream Mach number. Overall, both quantities depend weakly on $\be$, changing by $\lesssim$15\% when $\be$ increases from $0$ to $0.4$. \label{fig:xiav}}
\end{center}
\end{figure}

We now consider a field of incident acoustic waves. The response of the shock to incident acoustic perturbations can be expressed in terms of the root mean square (RMS) shock displacement and velocity, which are shown on the left and right panels of Figure~\ref{fig:xiav} as a function of nuclear dissociation parameter $\be$ for various values of ${\cal M}_1$. For fiducial mean flow parameters (${\cal M}_1 = 5$ and $\be = 0.2$), the shock displacement is $\simeq$0.58$\lambda_1 \langle A_p^2 \rangle^{0.5} $, where $\lambda_1$ and $A_p$ are the wavelength and amplitude of the incoming sound waves. The shock displacement has a weak dependence on $\be$, changing by less than $\sim$2\% when $\be$ grows from $0$ to $0.4$. The dependence on ${\cal M}_1$ is also weak for ${\cal M}_1 \gtrsim 5$, which is representative of the conditions in CCSNe. For example, the RMS displacement changes by only $\sim$10\% when the pre-shock Mach number ${\cal M}_1$ increases from $5$ to $100$. The RMS shock displacement velocity is shown on the right panel of Figure~\ref{fig:xiav} as a function of $\be$ for the same values of ${\cal M}_1$. The RMS displacement velocity is $\simeq$0.19$U_1 \langle A_p^2 \rangle^{0.5}$ for the fiducial mean flow parameters (${\cal M}_1 = 5$ and $\be = 0.2$). Similarly to the shock displacement, the shock velocity also has a weak dependence on $\be$, decreasing by $\sim$10\% when $\be$ increases from $0$ to $0.4$. The dependence on ${\cal M}_1$ also becomes weak for ${\cal M}_1 \gtrsim 5$. For example, when ${\cal M}_1$ increases from $5$ to $100$, the RMS shock velocity decreases by less than $\sim 10\%$. 

\begin{figure}[H]
\begin{center}
\includegraphics[angle=0,width=0.5\columnwidth, clip=false]{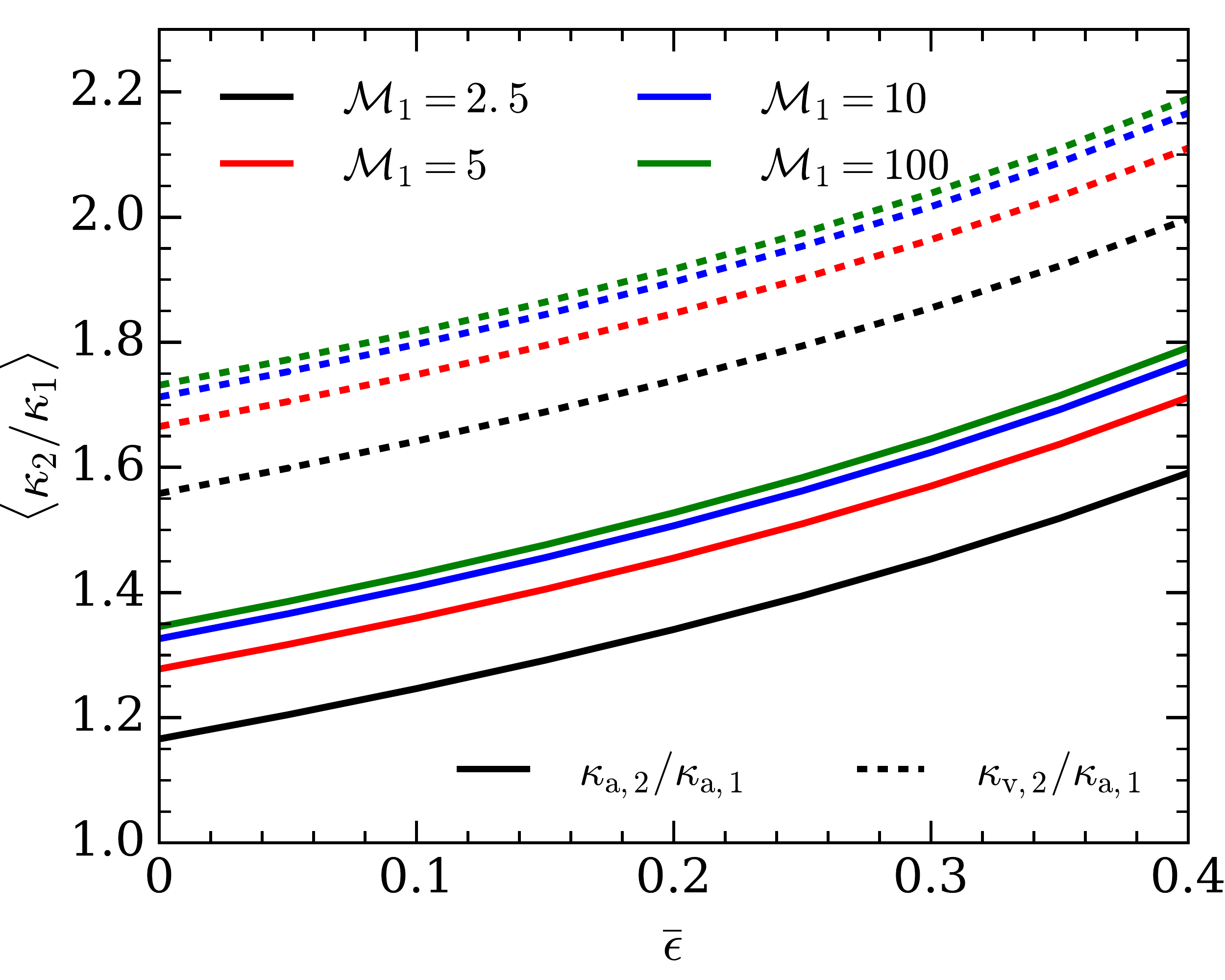}
  \caption{The ratio of the wavenumbers of acoustic waves (solid lines) and vorticity waves (dashed lines) in the post-shock region to that of incoming acoustic waves as a function of the nuclear dissociation parameter for various values of upstream Mach numbers ranging from $2.5$ to $100$. In all cases, the ratio grows with increasing $\be$ and ${\cal M}_1$ due to larger compression. The growth saturates beyond ${\cal M}_1 \sim 5$, reflecting the fact that shock wave parameters have weak dependence on ${\cal M}_1$ for ${\cal M}_1 \gtrsim 5$. For our fiducial parameters ($\be=0.2$ and ${\cal M}_1=5$), the ratio of the wavenumbers of the acoustic waves is $1.4$, while the ratio of incident acoustic waves to that of post-shock vorticity waves is $1.85$. These values are similar to the ratio of the wavenumbers of downstream and incident vorticity waves considered in \cite{abdikamalov:16}.
  \label{fig:k2k1}}
\end{center}
\end{figure}

When acoustic waves cross the shock, they undergo shock compression. The solid lines in Figure~\ref{fig:k2k1} show the ratios of the angle-averaged wavenumbers of post-shock acoustic waves to that of incident acoustic waves as a function of $\be$ for four values of ${\cal M}_1$ ranging from $2.5$ to $100$ (here, the angular  averaging is performed over in the propagative regime, i.e., $\psi_1 < \psi_\mathrm{cl}$ or $\psi_1 > \psi_\mathrm{cu}$). Since the shock compression grows with either $\bar\epsilon$ or ${\cal M}_1$, so does the ratio of wavenumbers. For our fiducial mean flow parameters ($\be=0.2$ and ${\cal M}_1=5$), the ratio equals $1.46$. The ratio ranges from $\sim$1.17 for weak shocks with small nuclear dissociation (${\cal M}_1\lesssim 2.5$ and $\be \lesssim 0.1$) to $\sim$1.8 for strong shocks with significant nuclear dissociations (${\cal M}_1\gtrsim 5$ and $\be \gtrsim 0.2$). The latter regime is more representative of the accretion shock flow parameters in CCSNe.

It is also instructive to compare the wavenumbers of vorticity waves generated in the post-shock region to that of incident acoustic waves, which is shown with dashed lines in Figure~\ref{fig:k2k1} as a function of $\be$ for the same four values of ${\cal M}_1$. For fiducial mean flow parameters ($\be=0.2$ and ${\cal M}_1=5$), the ratio is $1.85$. Overall, the ratio ranges from $\sim$1.56 for weak shocks (${\cal M}_1\lesssim 2.5$) with small nuclear dissociation ($\be \lesssim 0.1$) to $\sim$2.2 for strong shock with strong nuclear dissociations (${\cal M}_1\gtrsim 5$ and $\be \gtrsim 0.2$). These~values are similar to the ratios of the wavenumbers of downstream and upstream vorticity waves considered in \cite{abdikamalov:16}.

\begin{figure}[H]
\begin{center}
\includegraphics[angle=0,width=0.45\columnwidth, clip=false]{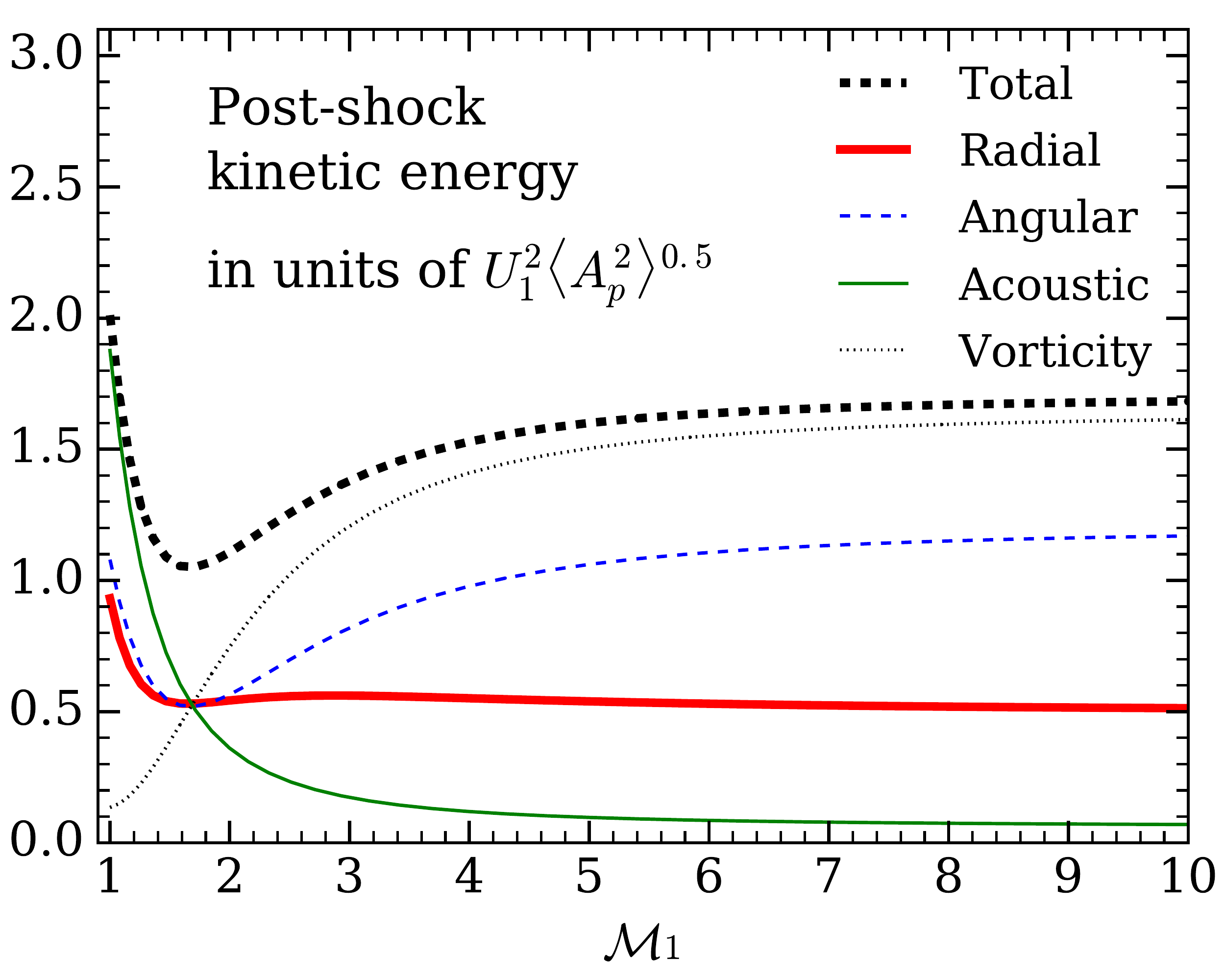}
\includegraphics[angle=0,width=0.45\columnwidth, clip=false]{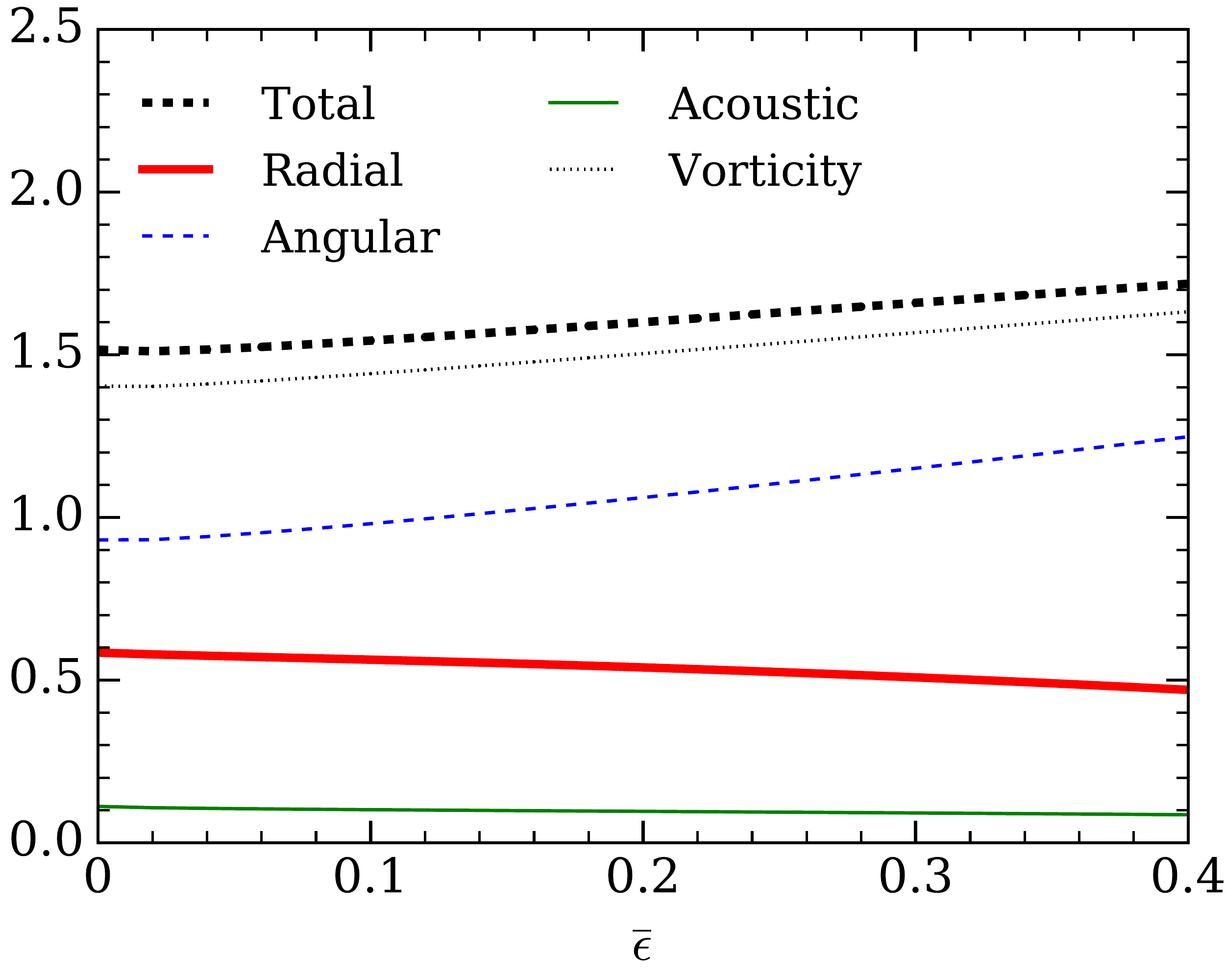}
  \caption{({\bf left panel}): The kinetic energy of post-shock perturbations as functions of upstream Mach number ${\cal M}_1$ for $\be=0.2$. In region $1 \lesssim {\cal M}_1 \lesssim 5$, the kinetic energies of various perturbations undergo rapid variations with ${\cal M}_1$. For larger ${\cal M}_1$, which is more representative of the situation in CCSNe, they exhibit almost no change with ${\cal M}_1$. In this regime, the kinetic energy of vorticity waves dominates the kinetic energy in the post-shock region, accounting for $94\%$ of the total kinetic energy of all perturbations at ${\cal M}_1 = 5$. The acoustic perturbations represent the remaining $4\%$ of the kinetic energy; ({\bf right panel}): the kinetic energy of post-shock perturbations as functions of nuclear dissociation parameter $\be$ for ${\cal M}_1=5$. The dependence of kinetic energy on $\be$ is somewhat weak and the perturbation energies change by at most $\sim 10\%$ when $\be$ increases from $0$ to $0.4$. While the total kinetic energy and angular kinetic energy increases with $\be$, the opposite happens to the radial kinetic energy in the post-shock region. \label{fig:kin_en}}
\end{center}
\end{figure}

The left panel of Figure~\ref{fig:kin_en} shows the specific kinetic energy of downstream acoustic and vorticity waves as a function of the upstream Mach number ${\cal M}_1$ for $\be=0.2$. For ${\cal M}_1 < 5$, the kinetic energies undergo rapid variations with ${\cal M}_1$, but for larger values of ${\cal M}_1$ (${\cal M}_1 \gtrsim 5$), which is representative of the conditions in CCSNe, these variations quickly subside and the kinetic energies slowly approach their asymptotic (${\cal M}_1 \rightarrow \infty$) values. For fiducial mean flow parameters (${\cal M}_1 = 5$ and $\be=0.2$), the~total kinetic energy in the post-shock region is $\simeq$1.6$U_1^2 \langle A_p^2 \rangle $.  In this regime, the vorticity waves contain $94\%$ of the total kinetic energy of velocity fluctuations. The remaining fraction is contained in acoustic waves. For incident vorticity or entropy wave perturbations, which we considered in \cite{abdikamalov:16,huete:18}, the~vorticity waves similarly provide the dominant contribution to the kinetic energy in the post-shock region \cite{abdikamalov:16,huete:18}.

The right panel of Figure~\ref{fig:kin_en} shows the specific kinetic energy of downstream acoustic and vorticity waves as a function of the nuclear dissociation parameter $\be$ for the upstream Mach number ${\cal M}_1=5$. Overall, we observe somewhat weak dependence on $\be$. The total kinetic energy increases from $1.50$ at $\be=0$ to $1.72$ at $\be=0.4$. For the same range of $\be$, the angular component of the kinetic energy increases from $0.93$ to $1.25$. In contrast, the radial component decreases from $0.59$ to $0.47$. Such an opposite dependence on $\be$ was also observed for the post-shock kinetic energy triggered by incoming vorticity waves \cite{abdikamalov:16}. 

\begin{figure}[H]
\begin{center}
\includegraphics[angle=0,width=0.45\columnwidth, clip=false]{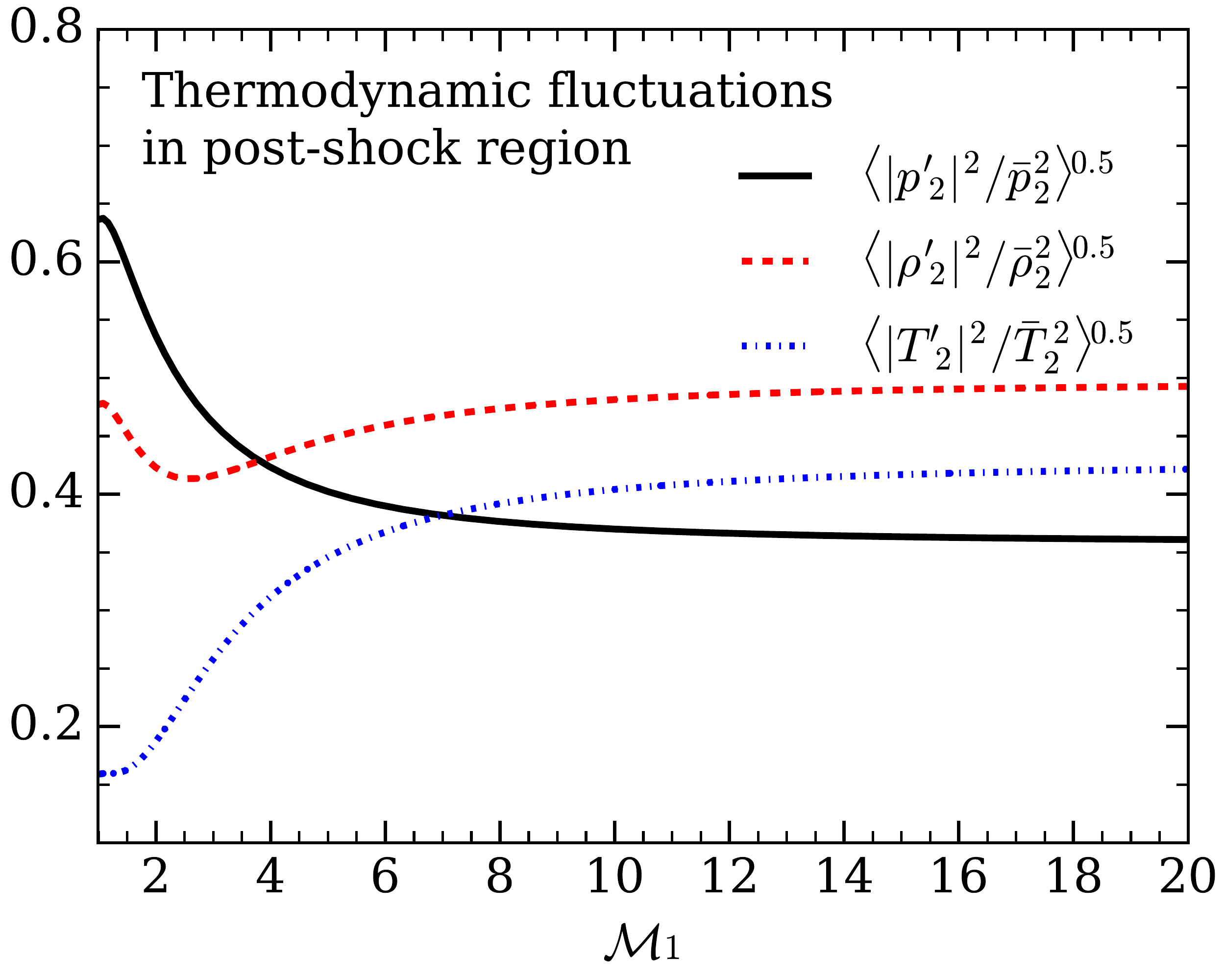}
\includegraphics[angle=0,width=0.45\columnwidth, clip=false]{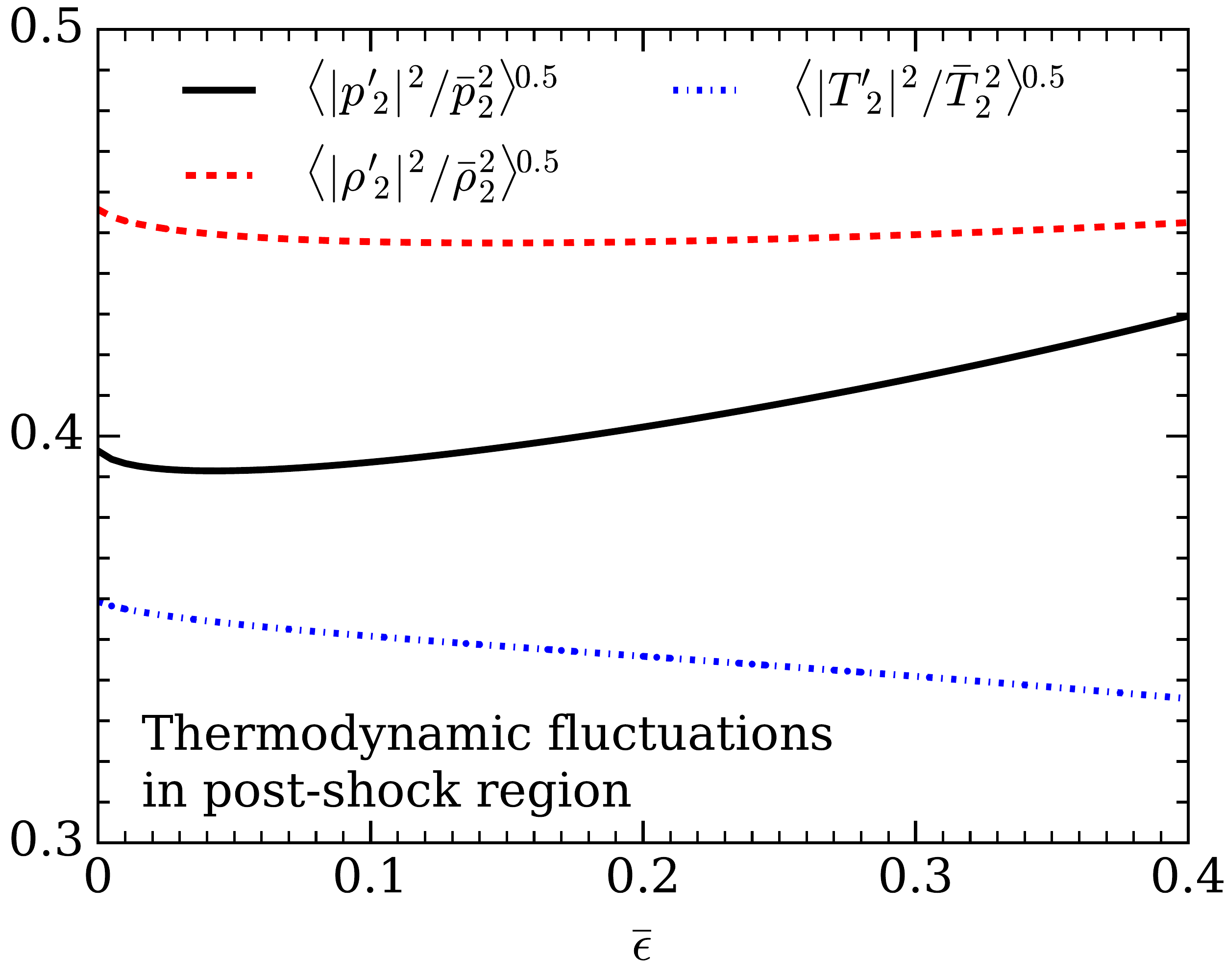}
  \caption{\textls[-15]{({\bf left panel}): variations of pressure (solid black line), density (dashed red line), and~temperature} (dashed dotted blue line) in the post-shock region as a function of upstream Mach number for $\be=0.2$ in units of $\langle A_p^2 \rangle^{0.5}$; ({\bf right panel}): variations of pressure (solid black line), density (dashed red line), and temperature (dashed dotted blue line) in the post-shock region as a function of nuclear dissociation parameter $\be$ for upstream Mach number of ${\cal M}_1=5$.\label{fig:prhoT}}
\end{center}
\end{figure}

The relative strengths of acoustic, vorticity, and entropy perturbations in the post-shock region are also reflected in the amplitudes of the variations of pressure, velocity, and density, respectively. The~left panel of Figure~\ref{fig:prhoT} shows the RMS fluctuations of these quantities as a function of upstream Mach number for $\be=0.2$. Overall, all the RMS fluctuations lie between $\sim$0.3 and $\sim$0.5 for ${\cal M}_1\gtrsim 5$. Similarly to post-shock kinetic energies, the dependence on ${\cal M}_1$ becomes weak for ${\cal M}_1 \gtrsim 5$. The~dependence of the same quantities on $\be$ at ${\cal M}_1=5$ is shown on the right panel of Figure~\ref{fig:prhoT}. The pressure fluctuations increase from $0.4$ to $0.43$ as $\be$ increases from $0$ to $0.4$. On the other hand, the temperature variations decrease from $0.36$ to $0.34$ for the same increase in $\be$. The density perturbations, which contain contributions from both acoustic and entropy waves, are somewhat insensitive to the value of $\be$, changing by less than $1\%$ as $\be$ increases from $0$ to $0.4$. 

\begin{figure}[H]
\begin{center}
\includegraphics[angle=0,width=0.49\columnwidth, clip=false]{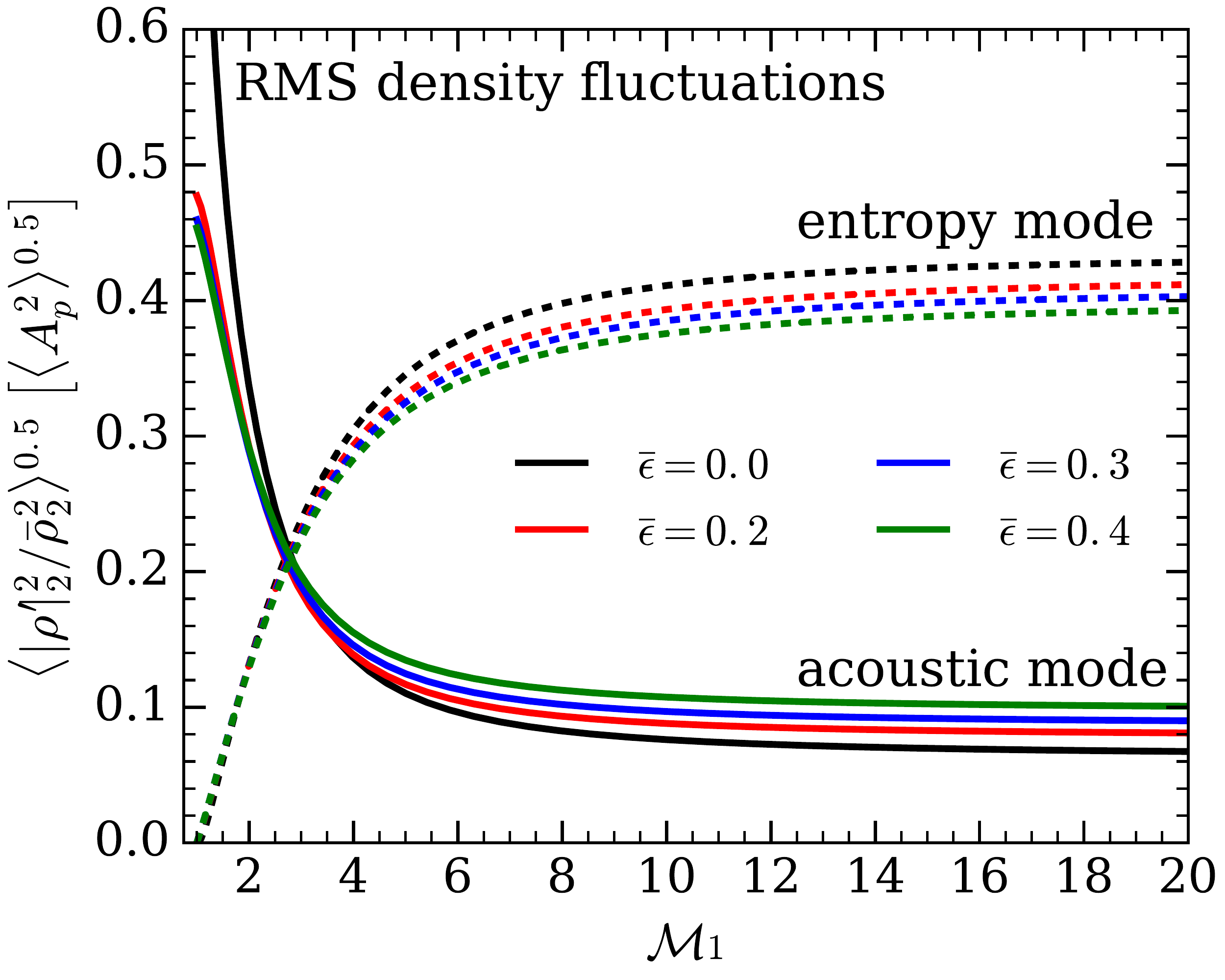}
  \caption{RMS density variations in the post-shock region as a function of upstream Mach number for different values of nuclear dissociation parameters $\be$. The solid lines represent the contribution of acoustic waves, while the dashed lines show the contribution of entropy modes in the post-shock region. In the weak shock limit (${\cal M}_1 \simeq 1$), most of the contribution to density fluctuations are due to acoustic waves. The contribution of entropy modes grows with ${\cal M}_1$, while the opposite happens to the contribution of acoustic waves.\label{fig:rhoe}}
\end{center}
\end{figure}

Figure~\ref{fig:rhoe} shows the contribution of acoustic (solid lines) and entropy waves (dashed lines) to the RMS density variations as a function of upstream Mach number ${\cal M}_1$ for various values of $\be$ in units of $\langle A_p^2 \rangle^{0.5}$. In the limit of small ${\cal M}_1$ ($ {\cal M}_1 \simeq 1$), most of the contribution comes from acoustic waves, while the contribution of entropic modes is negligible. This is not surprising since, in the limit ${\cal M}_1\simeq 1$, acoustic waves cross the shock without perturbing it, which prevents generation of entropy waves in the post-shock region. However, the contribution of the acoustic waves decreases fast with increasing ${\cal M}_1$, while the opposite happens to the contribution of entropy modes. For ${\cal M}_1 \gtrsim 5$, which represents the regime in CCSNe, most of the density variations in the post-shock region are due to the entropy waves, contributing more than $\sim$70\% of the total density variations. In this case, the amplitude of density fluctuations due to entropy modes is $\sim$0.33$\langle A_p^2 \rangle^{0.5}$. 

\subsection{Implications for CCSN Explosion Condition}
\label{sec:implications}

Generated by the acoustic perturbations of the shock, the entropy waves in the post-shock region become buoyant and drive additional convection, creating a more favorable condition for producing explosion. An order-of-magnitude estimate for the the reduction of the critical neutrino luminosity for producing explosion can be inferred using the model of \cite{mueller:16}:
\begin{equation}
\label{eq:cl00}
\frac{\Delta L_\mathrm{crit}}{L} \simeq \frac{0.15 \pi}{\ell \eta_\mathrm{acc} \eta_\mathrm{heat}} \frac{\langle |\rho_2|'^2 \rangle^{0.5} }{\bar \rho_2}.
\end{equation}

Here, $\ell$ is the angular wavenumber of the dominant mode of perturbation and $\sqrt{\langle |\rho_2|'^2 \rangle}$ is the RMS density perturbations \cite{huete:18}. $\eta_\mathrm{acc}$ and $\eta_\mathrm{heat}$ are the efficiencies of accretion and neutrino heating. Using typical values $\eta_\mathrm{acc}=2$ and $\eta_\mathrm{heat}=0.1$ \cite{mueller:17},
\begin{equation}
\label{eq:cl0}
\frac{\Delta L_\mathrm{crit}}{L} \simeq \frac{2.36}{\ell} \frac{\langle |\rho_2|'^2 \rangle^{0.5} }{\bar \rho_2}
\end{equation}
for typical problem parameters.  As we can see in Figure~\ref{fig:rhoe}, $\langle |\rho_2|'^2 \rangle^{0.5}/{\bar \rho_2} \simeq 0.33 \langle A_p^2 \rangle^{0.5}$ for fiducial mean flow problem parameter and it has a weak dependence on $\be$ and ${\cal M}_1$ for ${\cal M}_1 \gtrsim 5$. Substituting this into Equation (\ref{eq:cl0}), we obtain estimate 
\begin{equation}
\label{eq:cl1}
\frac{\Delta L_\mathrm{crit}}{L} \simeq \frac{0.78}{\ell} \langle A_p^2 \rangle^{0.5}. 
\end{equation}

For $\ell=2$ perturbations, which has the largest impact on the explosion condition \cite{mueller:15}, we get
\begin{equation}
\label{eq:cl2}
\frac{\Delta L_\mathrm{crit}}{L} \simeq 0.39 \langle A_p^2 \rangle^{0.5}. 
\end{equation}

For $\langle A_p^2 \rangle^{0.5} \sim 0.13$, which is close to the upper limit, this leads to a $\sim$5.2\% decrease in the critical luminosity. This is significantly smaller than the effect of vorticity waves, which can be as high as $\sim$24\% \cite{huete:18}. This suggests that upstream acoustic perturbations play a less important role compared to vorticity perturbations. This result seems to be at least consistent with the results of 3D neutrino-hydrodynamics simulations of \cite{mueller:17}, who studied the impact of convective perturbations in the O burning shell. Since acoustic waves travel with respect to the flow, they must arrive at the shock earlier than the O burning shell. However, the evolution of the shock is affected significantly only after the O shell arrives at the shock, suggesting that the acoustic waves may have a modest impact on the shock dynamics. However, Equation~(\ref{eq:cl2}) is based on a number of approximate assumptions and it is meant to provide only an order-of-magnitude estimate for the reduction of the critical luminosity (e.g., \cite{huete:18}). A more rigorous estimate will be provided in a future work.

\section{Conclusions}
\label{sec:summary}

In this paper, we studied the impact of acoustic waves generated by convective instabilities in nuclear burning shells of core-collapse supernova (CCSN) progenitors. Using a linear perturbation theory, we analyzed the interaction of these waves with the supernova shock.~We modeled the unperturbed flow as a one-dimensional uniform flow that is compressed by a planar shock perpendicular to the flow. We calculated the properties of vorticity, entropy, and acoustic waves generated in the post-shock region as a result of the perturbation of the supernova shock by upstream acoustic perturbations. The dissociation energy is assumed to be invariant to perturbations. Despite all the upstream kinetic energy of the velocity perturbations being of acoustic type, we find that the kinetic energy in the post-shock region is dominantly of rotational type, due to the shock-generated vorticity. Correspondingly, acoustic waves in the post-shock flow account for a tiny fraction of the kinetic energy. For example, for our fiducial mean flow parameter (${\cal M}_1=5$ and $\be=0.2$), the contribution of vorticity waves is $94\%$, while the remaining $6\%$ is due to acoustic waves. The density perturbations in the post-shock region are mostly due to entropy waves, contributing $74\%$ of the total value for our fiducial mean flow parameters. Once in the post-shock region, these entropic perturbations become buoyant and generate additional turbulence in the gain region. Using the model of M{\"u}ller et al \cite{mueller:16}, we~estimated the reduction of the critical neutrino luminosity necessary for producing explosion to be $\lesssim$5\%. \linebreak This~is about 3--5 times smaller than the effect of incident vorticity waves, which tentatively suggests that the incoming acoustic waves are likely to have a relatively modest impact on the explosion mechanism of CCSNe. However, this estimate is based on the number of assumptions and approximations. In~particular, we assumed isotropic distribution of incoming acoustic waves and it is unclear how good this approximation is. By the time the incoming acoustic waves encounter the supernova shock, the post-shock region is likely to have a fully developed neutrino-driven convection and/or standing accretion shock instability. The interaction of these instabilities with the fluctuations generated by the incoming perturbations needs to be described within a more rigorous theory (e.g., \cite{takahashi:16,mabanta:18}). This will be the subject of future studies.

\vspace{6pt} 

\acknowledgments{We thank Thierry Foglizzo, Bernhard M\"uller, and David Radice for fruitful discussions. Funding from Nazarbayev University ORAU grant SST 2015021 and Ministry of Education and Science of the Republic of Kazakhstan state-targeted program BR05236454 is acknowledged.}

\authorcontributions{Ernazar Abdikamalov initiated the project, contributed to the development of the methodology, analysis of the results, and preparation of the manuscript. C\'esar Huete contributed to the development of the formalism, including the aspects related to the treatment of the nuclear dissociation energy. He also helped with the analysis of the results and preparation of the manuscript. Shapagat Berdibek and Ayan Nussupbekov helped with the derivation of the equations, analysis of the results, and preparation of the manuscript.} 

\conflictsofinterest{The authors declare no conflict of interest.} 



\reftitle{References}

\end{document}